\documentclass[%
 reprint,
superscriptaddress,
 amsmath,amssymb,
 aps,
prl
]{revtex4-1}

\usepackage{placeins}
\usepackage{subcaption}
\usepackage{graphicx}
\usepackage{dcolumn}
\usepackage{bm}

\begin{document}

\title{Radio Frequency Magneto-Optical Trapping of CaF with High Density}

\author{Lo\"ic Anderegg}
\email{anderegg@g.harvard.edu} 
\author{Benjamin Augenbraun}
\affiliation{Department of Physics, Harvard University, Cambridge, MA 02138, USA}
\affiliation{Harvard-MIT Center for Ultracold Atoms, Cambridge, MA 02138, USA}

\author{Eunmi Chae}
\altaffiliation[Present address: ]{Photon Science Center, School of Engineering, the University of Tokyo, Japan 113-8656}
\author{Boerge Hemmerling}
\altaffiliation[Present address: ]{Department of Physics, University of California, Berkeley, California 94720, USA}
\author{Nicholas R. Hutzler}
\author{Aakash Ravi}
\affiliation{Department of Physics, Harvard University, Cambridge, MA 02138, USA}
\affiliation{Harvard-MIT Center for Ultracold Atoms, Cambridge, MA 02138, USA}

\author{Alejandra Collopy}
\author{Jun Ye}
\affiliation{JILA, National Institute of Standards and Technology and University of Colorado, Boulder, CO 80309, USA }

\author{Wolfgang Ketterle}
\affiliation{Harvard-MIT Center for Ultracold Atoms, Cambridge, MA 02138, USA}
\affiliation{Department of Physics, Massachusetts Institute of Technology, Cambridge, MA 02139, USA }

\author{John Doyle} 
\affiliation{Department of Physics, Harvard University, Cambridge, MA 02138, USA}
\affiliation{Harvard-MIT Center for Ultracold Atoms, Cambridge, MA 02138, USA}

\date{\today}

\begin{abstract}

We demonstrate significantly improved magneto-optical trapping of molecules using a very slow cryogenic beam source and RF modulated and DC magnetic fields. The RF MOT confines $1.1(3) \times 10^5$ CaF molecules at a density of $4(1) \times 10^6$~cm$^{-3}$, which is an order of magnitude greater than previous molecular MOTs. Near Doppler-limited temperatures of $340(20) \mu$K are attained. The achieved density enables future work to directly load optical tweezers and create optical arrays for quantum simulation. 


\end{abstract}

\maketitle


The field of ultracold molecules is rapidly expanding, pushed forward by favorable prospects for advances in quantum simulation~\cite{zoller06,carr09}, quantum information~\cite{demille02qi,yelin06}, quantum chemistry~\cite{krems08,ni10}, and precision measurements~\cite{ACME14,hinds12}. The rich internal structure of molecules, including rotational and vibrational modes, as well as their long-range and anisotropic dipolar interactions, make diatomic molecules ideal extensions of current work in these fields. While great progress has been made by assembling bi-alkalis~\cite{ni08,ospelkaus10,moses15}, direct cooling of molecules allows one to realize an increase in chemical diversity. This diversity is of particular interest in the case of $^2\Sigma$ molecules, where the unpaired electron spin would allow one to realize lattice spin models~\cite{zoller06,pupillo08} and the onset of topologically ordered states~\cite{zoller07}. 

The same internal degrees of freedom that make molecules so interesting for scientific applications add further complexity to the cooling and trapping process. Despite this, several groups have sought to cool  molecules using direct laser cooling with simultaneous application of multiple laser frequencies, in particular the construction of molecular MOTs. 
Magneto-optical traps (MOTs) have long been the workhorse of cold atom experiments, producing (sub-)Doppler-limited temperatures of many atomic species. Recently, following the first demonstration of 2D magneto-optical compression of YO~\cite{hummon13}, the first molecular MOT was created with SrF~\cite{barry14,norrgard16RF,steinecker16}. A major challenge in reaching this point was the very low capture velocity of the molecular MOT due to many internal molecular states and the resulting low photon scattering rate. To extend the MOT to other species and improve the number of trapped molecules, laser slowing of YO~\cite{yeo15} and CaF~\cite{cafslow16,tarbutt17cafslow}, RF magneto-optical compression of CaF~\cite{chae17}, and, very recently, a MOT of CaF based on static magnetic fields (DC MOT) were demonstrated~\cite{truppe17}. 

We demonstrate and characterize RF and DC MOTs of CaF and observe a sizable increase in both trapped number and density over previous molecular MOTs. 
By employing a slow two-stage buffer-gas beam source for more efficient MOT loading, the RF MOT traps  $1.1(3)\times 10^5$ CaF molecules at a density of $n_0=4(1) \times 10^6$~cm$^{-3}$. Our work indicates a general route to increased MOT densities and number, which are desired for several important applications.
%

\begin{figure*}
\centering
   \begin{subfigure}{0.33\textwidth}
   \includegraphics[scale=.38]{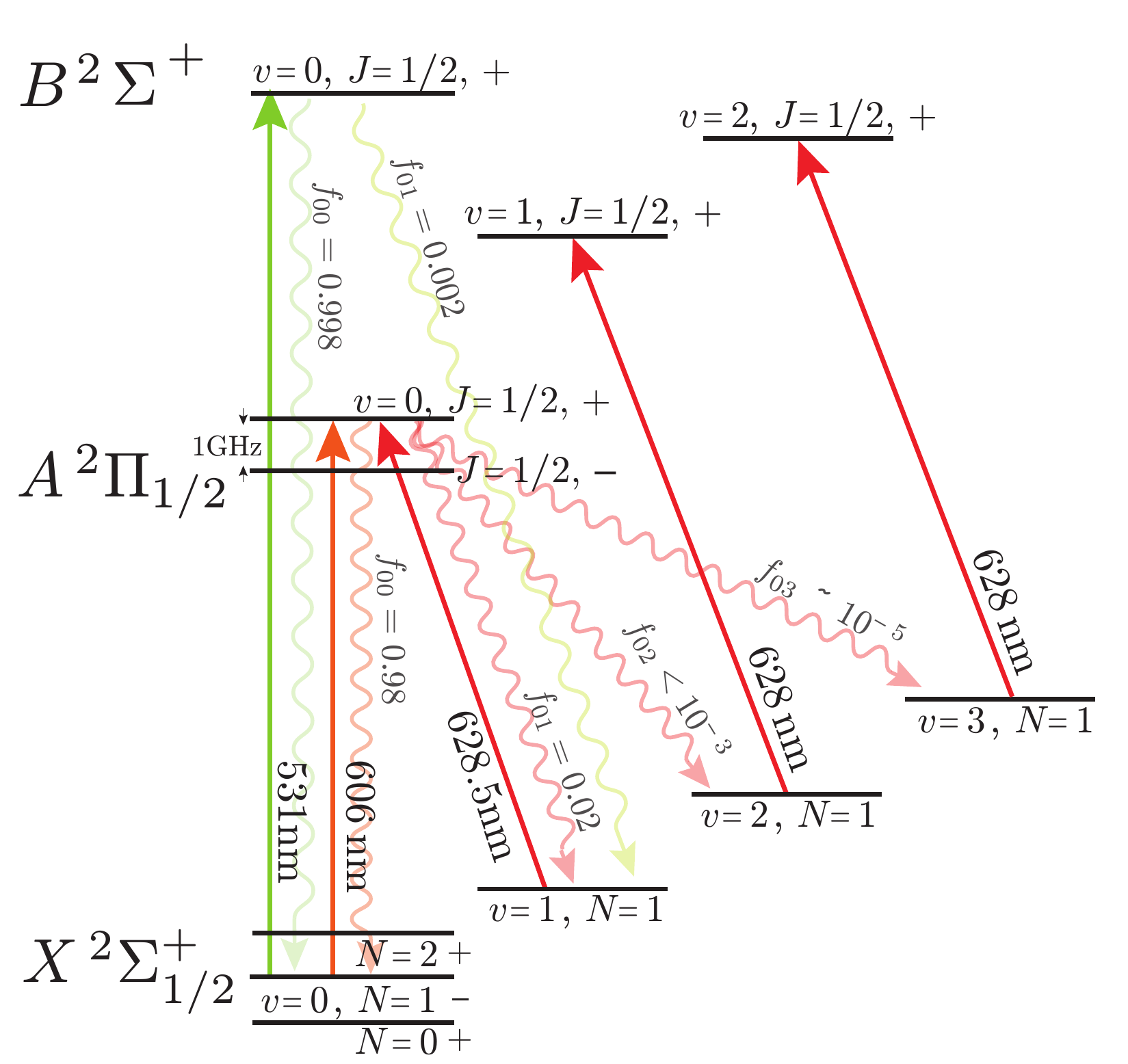} 
\caption{}
\label{fig:transitions}
   \end{subfigure}
     ~ 
       \begin{subfigure}{0.22\textwidth}
        \includegraphics[scale=.47]{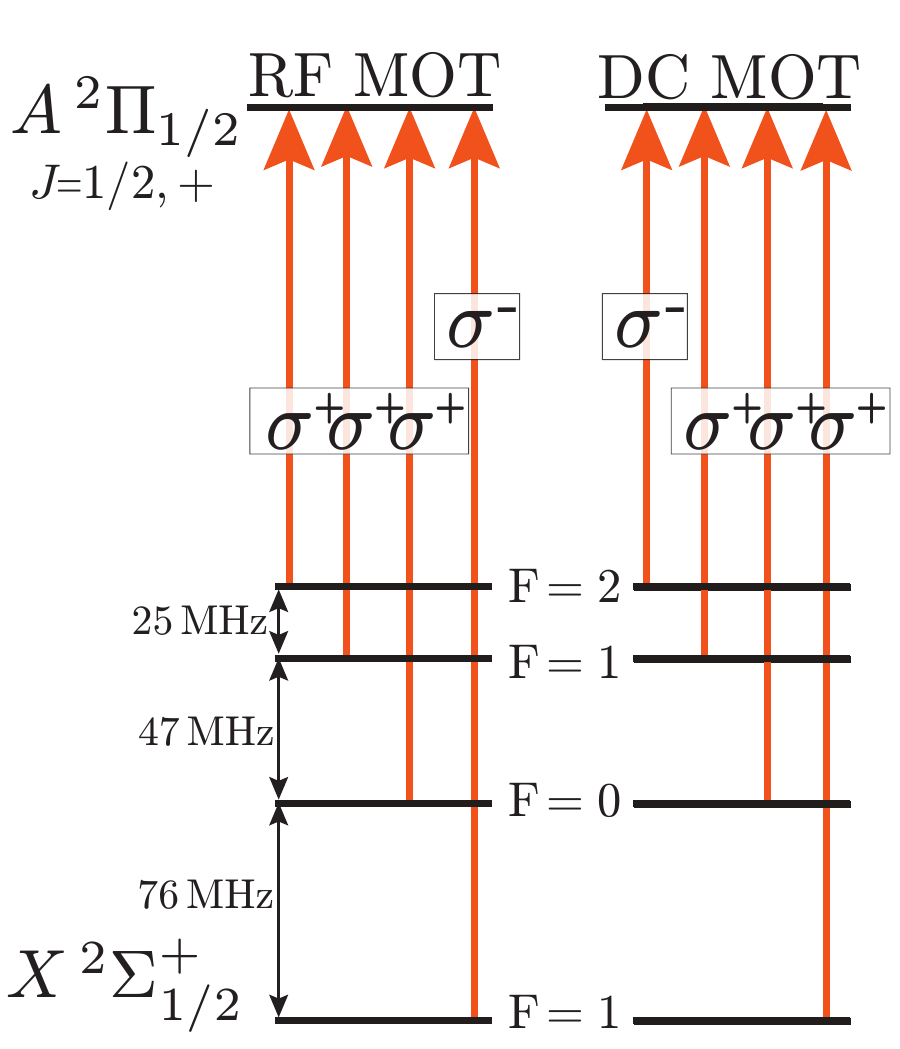} 
                \caption{}
        \label{fig:pol}  
   \end{subfigure}
    ~ 
       \begin{subfigure}{0.39\textwidth}
        \includegraphics[scale=.5]{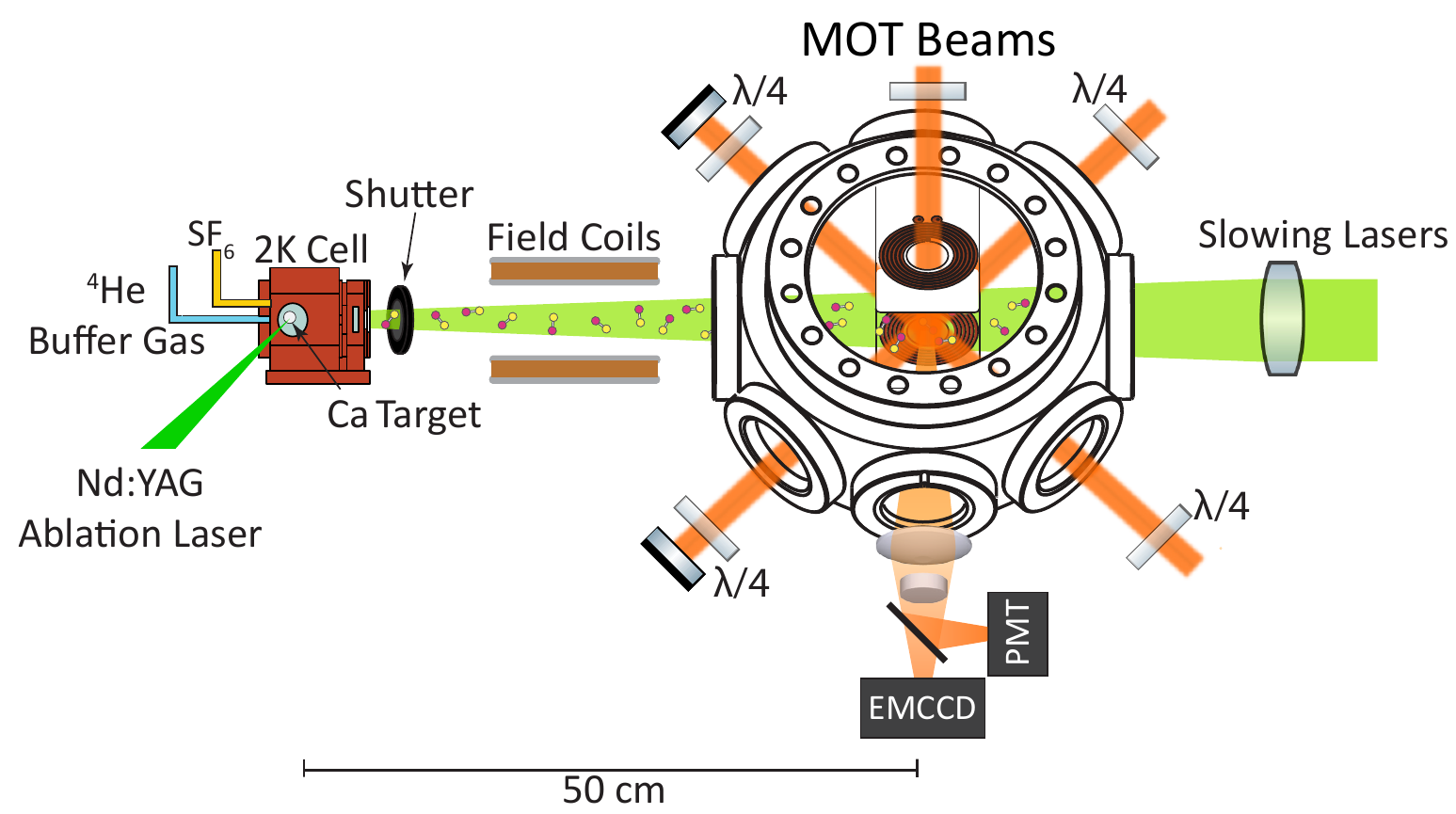} 
                \caption{}
   \end{subfigure}
      ~
\caption{(a)~Level diagram of CaF. Solid lines indicate laser excitation frequencies. The 531~nm X-B transition is used for slowing, while the 606~nm X-A transition is used for the MOT. Wavy lines indicate potential decay paths from the main photon cycling transitions to higher vibrational levels, with their corresponding Franck-Condon factors indicated. (b)~Polarization schemes used to drive the X-A MOT transitions. The hyperfine structure of the ground state is addressed with 3 AOMs with the proper frequency shifts. The hyperfine structure of the A state is unresolved. (c)~The experimental layout. } 
\end{figure*}

A typical atomic MOT operates by cycling photons on a $F \rightarrow F+1$ transition (often called a ``type-I" MOT), where $F$ is the total angular momentum. However, in diatomic molecules such as CaF, photons must be cycled on a $F \rightarrow F-1$ transition (a so-called ``type-II" MOT). The relevant energy levels  for CaF are shown in Figure~\ref{fig:transitions}. While this provides a rotationally closed transition for the MOT, it also leads to dark states that would result in no average trapping force~\cite{stuhl08}. To prevent this effect, one must remix dark and bright states. One option is to create an RF MOT, where both the optical polarizations and magnetic field are alternated at a timescale matching the pumping rate into these dark states~\cite{damop13,hummon13}. In some cases, a DC MOT can still be achieved due to remixing that occurs naturally in the multi-level structure of molecules. CaF offers a structure for this due to the spacing of the $F=1$ and $F=2$ hyperfine ground states, where one can exploit a ``dual frequency"~\cite{tarbutt15} effect to achieve strong trapping and cooling forces. CaF is favorable for magneto-optical trapping with a relatively low mass and diagonal Franck-Condon factors, $.98$ for X-A(606~nm) and $.999$ for X-B(531~nm), $\Gamma \sim 2 \pi \times 8$~MHz \cite{chae17}. With two lasers to repump the vibrational $v=1$ and $v=2$ states, $\sim4\times 10^5$ photons may be scattered on the X-A transition (a typical number needed for laser slowing and trapping). Just one vibrational repump laser is necessary to scatter  $\sim3\times 10^4$ on the X-B transition.

In this experiment, CaF molecules are produced by laser ablation inside of a two-stage cryogenic buffer-gas cell~\cite{lu11}. The cell is cooled to 1.5~K by pumping on a $^4$He bath attached to the cell. A solid Ca target is ablated with a 12~mJ, 10~ns pulsed second harmonic Nd:YAG laser. Sulfur hexafluoride (SF$_6$) is flowed into the cell~\cite{tarbutt17cafslow} where it reacts with the ablated Ca to produce CaF, which thermalizes with cold $^4$He gas flowed into the cell at a rate of 4~sccm. The CaF is cooled to roughly 2~K, at which temperature the molecular beam's thermal rotational distribution is peaked at the $N=1$ rotational state, needed for rotationally closed photon cycling. Use of a two-stage buffer-gas cell reduces the boosting effects as the molecules are extracted from the cell~\cite{lu11},  producing a CaF beam with a mean forward velocity $v_0\sim80$~m/s and a velocity spread of $\Delta v_0\sim$50~m/s, lower than the typical $v_0\sim150$~m/s velocities found in single-stage sources. The MOT chamber pressure is at $2 \times 10^{-8}$~Torr when the buffer gas flow is on, and rises slightly when the laser ablation takes place. To prevent the He buffer gas from limiting the MOT lifetime, a shutter is installed in vacuum between the output of the source and the MOT. It is opened for 3~ms following ablation, prior to molecule slowing.

In order to be captured by the MOT, the CaF molecules must be slowed to $\lesssim$5~m/s. This is accomplished with white-light slowing~\cite{shuman10,cafslow16}, where a counter-propagating, frequency-broadened laser beam applies radiation pressure to the molecules and decelerates them without falling out of resonance with the Doppler-shifted velocity distribution. The slowing light is tuned to the X-B transition in order to decouple the cycling transition from the vibrational repump transitions, improving the number of slow molecules from our previous work~\cite{cafslow16} by nearly an order of magnitude. A 500~mW X-B$(v=0,v^\prime=0)$ beam and a 50~mW X-A$(v=1,v^\prime=0)$ repump laser are combined on dichroic mirrors and passed through a high modulation index electro-optic modulator (EOM) at $f_0 = 4.5$~MHz to broaden the laser spectrum by $\sim$300~MHz. The combined laser beams counter-propagate against the molecular beam with a $1/e^2$ size of 20~mm at 1~m from the cryogenic cell, converging to 4~mm at the exit of the cell. This focusing applies a small transverse confining force to the molecular beam to reduce transverse pluming of the slowed molecules. A transverse magnetic field (15~G) is applied along the slowing distance to precess molecules from dark magnetic sublevels back into bright states. The slowing lasers are turned on 3.25~ms after the ablation laser fires and remain on for $\sim$10-14~ms. The short slowing distance required (due to the slow beam source and low mass, compared to SrF), allows the two stage source to be placed only 50~cm from the MOT region, increasing the solid angle of capturable molecules.

After the CaF is slowed, they enter the MOT region where the six MOT beams drive the X-A($v=0,1 \rightarrow v^\prime=0$) transitions. RF sidebands produced by a series of acousto-optic modulators (AOMs) address all hyperfine transitions. The polarization of each hyperfine component can be set independently, allowing the use of the optimal polarization scheme for both the DC~\cite{tarbutt15} and RF MOTs (Figure \ref{fig:pol}). At full power, each MOT beam contains 60~mW of X-A$(v=0,v^\prime=0)$ light and 55~mW of X-A$(v=1,v^\prime=0)$ light. The powers in each hyperfine sideband are balanced to within $20\%$. The MOT beams have a $1/e^2$ diameter of 11~mm. We define $I_0 \equiv 400$~mW/cm$^2$ ($s_0\sim1$) as the maximum intensity of the X-A$(v=0,v^\prime=0)$ transition at the center of the MOT. The X-A$(v=2,v^\prime=1)$ repump laser beam enters into the chamber through the slowing laser window.

The MOT coils consists of two pairs of OFHC copper spirals, each mounted on each side of two alumina boards. An axial field gradient of 25~G/cm is found to produce the largest MOT signal for the DC MOT. To produce the field gradient necessary for the RF MOT, we use two RF amplifiers each producing ~200~W (one for each pair of coils), resulting in a field gradient of 14~G/cm RMS. The use of symmetric amplifiers nearly eliminates any RF electric fields, which would mix opposite parity states of the $\Lambda$-doubled A-state, $J=1/2(+)$ and $J=1/2(-)$ (Figure \ref{fig:transitions}). This prevents unwanted decay to the dark X$(N=0,2)$ rotational states. Previous RF molecular MOTs~\cite{norrgard16RF} required microwaves to remix these dark states, which reduced the overall scattering rate due to coupling of additional ground states.

\begin{figure}
\begin{center}
\includegraphics[height=2.3in]{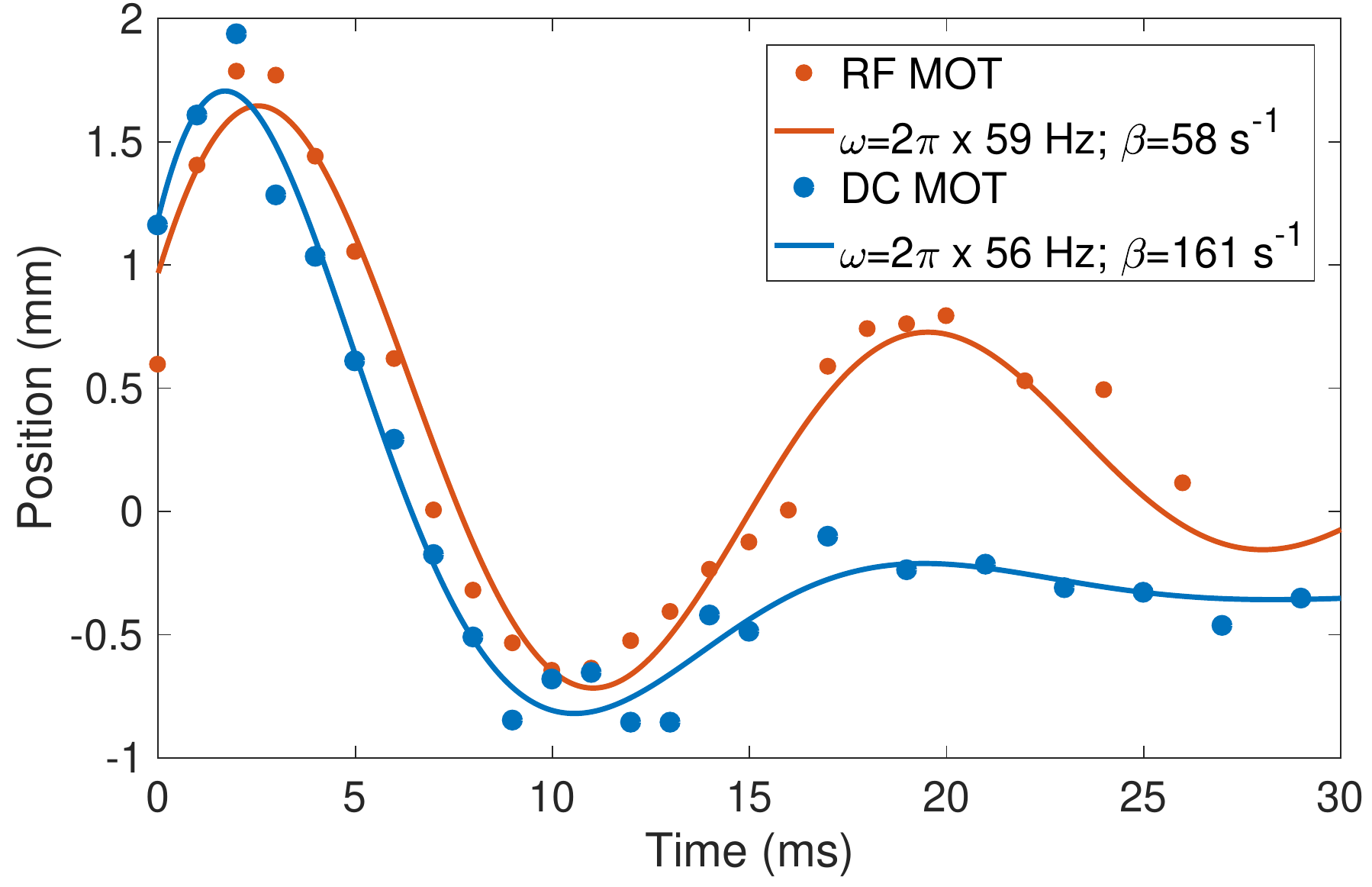} 
\caption{RF and DC MOT oscillations following a 1~ms push with the slowing laser. The equilibrium positions of the RF and DC MOTs are different since the DC remixing field in the slowing region shifts the center position of the DC MOT.}
\label{fig:spring}
\end{center}
\end{figure}

\begin{figure}
        \centering
        \includegraphics[height=2.8in]{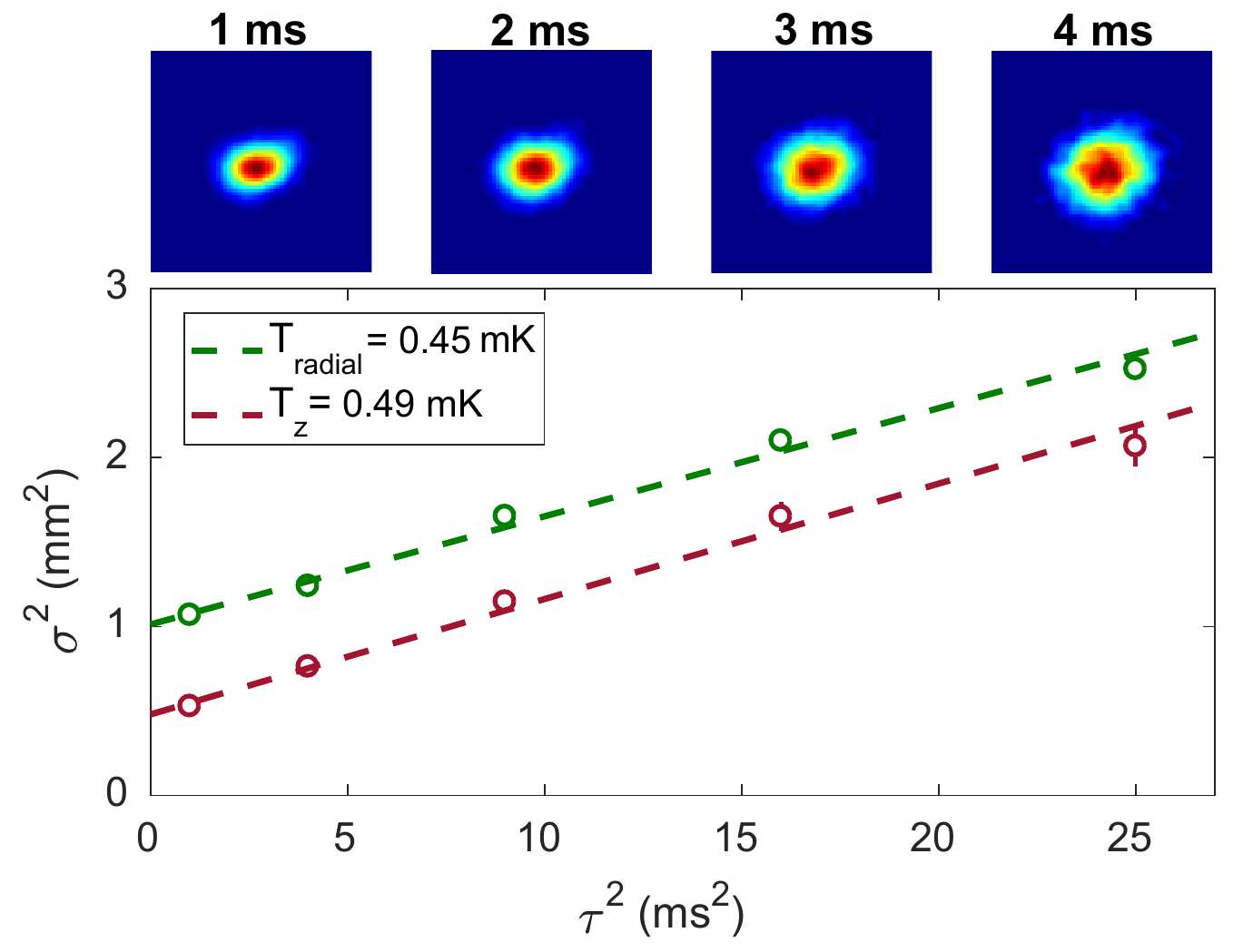}
        \caption{Time-of-flight expansion of the molecular cloud to measure temperature, shown here for the RF MOT after ramping down to $I_0/32$. Images are captured on an EMCCD with 1ms exposure and 15 averages. Image field of view is 10~mm $\times$ 10~mm. The dashes lines are fits to the TOF model (see text). We find nearly equal temperatures for the two dimensions, with the difference in the initial cloud size expected from the gradient produced in anti-Helmholtz coils.}
\label{temp}
\end{figure}

The molecules in the trap are detected via their MOT fluorescence (X-A transition at 606~nm), which is simultaneously recorded on both a PMT and an EMCCD for time and spatial information, respectively. A multi-lens objective in front of both devices is used to image and spatially filter the MOT fluorescence. Custom self-glued UHV windows~\cite{barrythesis} and blackening of the chamber are used to achieve a low background light scattering rates. The peak MOT intensity occurs at 20~ms after the ablation laser, and images are typically recorded at $t > 40$~ms. Typical MOT images after TOF measurements are shown in Figure \ref{temp}.

Scanning the frequency of the X-A$(v=0,v^\prime=0)$ transition showed a maximum MOT fluorescence signal at a detuning $\Delta = -8$~MHz from the transition center frequency and a FWHM of $\sim8$~MHz. Similar scans of both the first and second repump lasers showed peaks at $\Delta = 0$~MHz from their centers, with FWHMs of 30 and 40~MHz respectively.

\begin{figure*}
    \centering
    \begin{subfigure}{0.49\textwidth}
        \centering
        \includegraphics[height=2.5in]{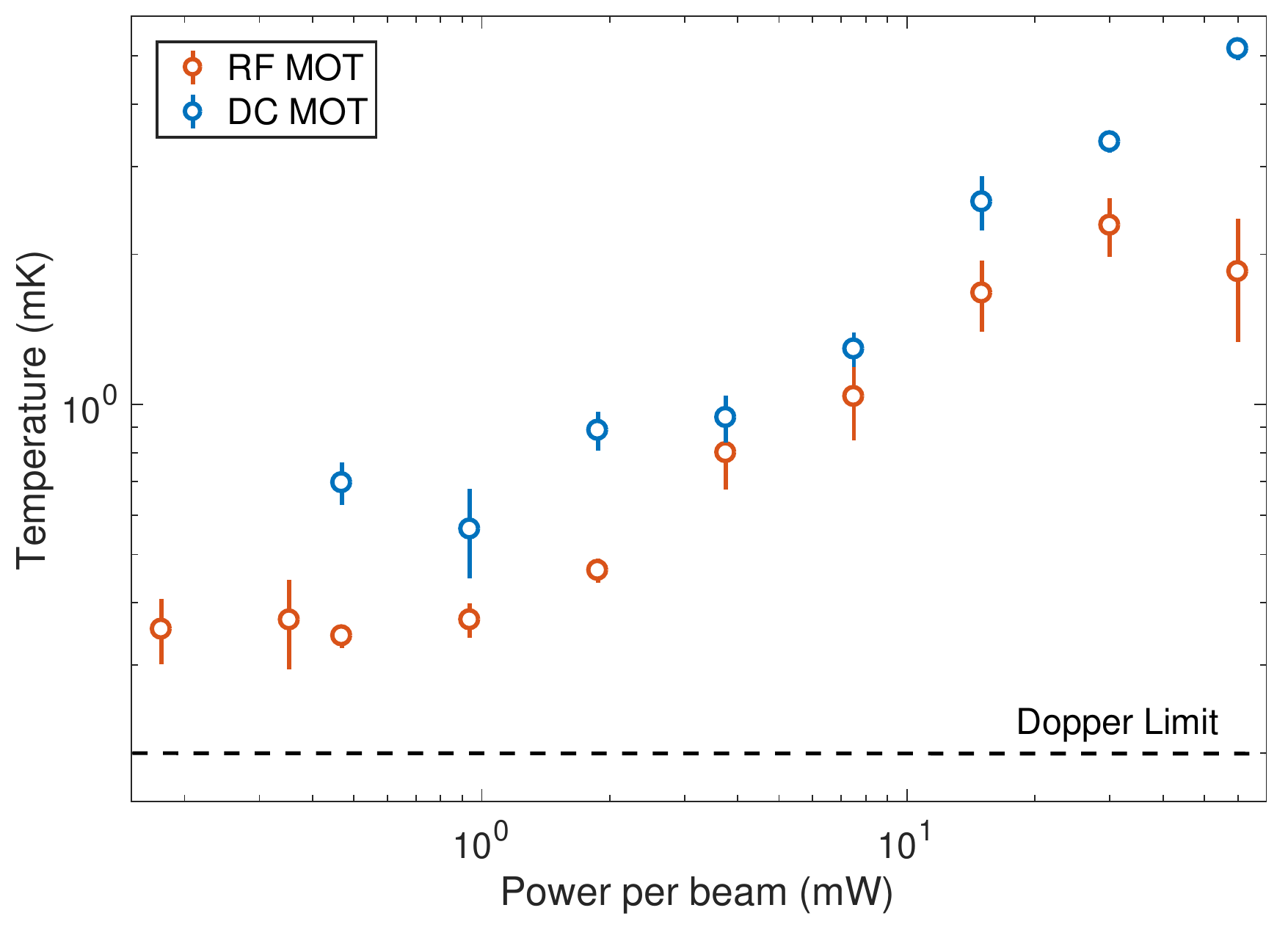}
        \caption{}
    \end{subfigure}%
    ~
    \begin{subfigure}{0.49\textwidth}
        \centering
        \includegraphics[height=2.5in]{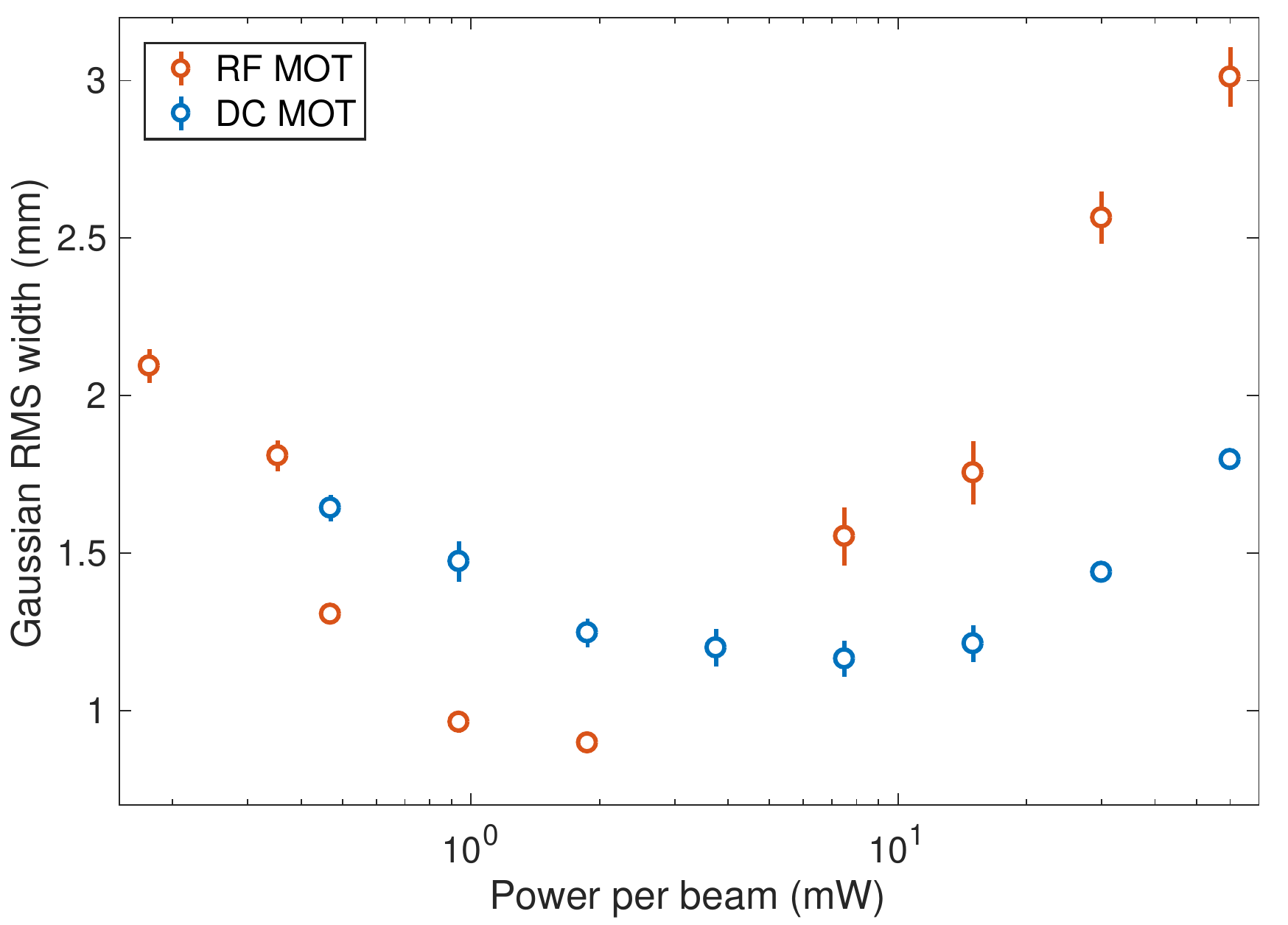}
        \caption{}
    \end{subfigure}
\caption{(a) MOT temperature vs MOT beam intensity following 15ms ramp and 10ms holding. The temperature is defined as $T_{axial}^{1/3}\times T_{radial}^{2/3}$ (b) MOT size (Gaussian RMS width) following the ramp and hold.}
\label{tempscan}    
\end{figure*}

The MOT lifetime is determined by measuring the decay of the fluorescence on the PMT. Decreasing the intensity of the MOT beams, and thus the scattering rate, increases the MOT lifetime inversely proportionally to the scattering rate. We find a MOT lifetime of $\tau=20$~ms without repumping the $v=3$ state, in line with expectations based on the CaF's Franck-Condon factors. Adding a $v=3$ repump laser, we find a lifetime of $\tau=85$~ms. We find that DC and RF MOT have roughly equal lifetimes at the same scattering rate, confirming the electric field in the RF MOT is suppressed to a level less than 1/10 of the decay rate into $v=3$, as expected from the use of the symmetrical amplifier configuration.

The trap frequency and damping constant are determined by recording the position of the molecular cloud following a 1~ms pulse with the slowing beam (Figure \ref{fig:spring}). At a gradient of 14~G/cm and the MOT beams at an intensity of $I_0$, the molecular cloud in the DC MOT oscillates in an underdamped fashion at $2\pi \times 56(2)$~Hz with a damping coefficient of $161(20)$~s$^{-1}$. The RF MOT with a 14~G/cm RMS field gradient oscillates at $2\pi \times 59(2)$~Hz and a damping coefficient of $58(10)$~s$^{-1}$.

The temperature of the MOT is determined by time-of-flight (TOF) expansion (Figure \ref{temp}). The MOT beams and magnetic field are turned off for varying lengths of time, $\tau_{TOF}$, followed by a 1 ms MOT laser imaging pulse. A laser intensity of $I_0/32$ is used to reduce any heating effects during imaging. This also ensures we image the MOT below saturation. We fit the radial and axial temperatures to a 2D Gaussian model with the difference of initial and final width after expansion, $\sigma^2 -\sigma_0^2 = k_B T \tau_{TOF}^2/m $. To achieve lower temperatures, while maximizing the number of trapped atoms, we load our MOT at full intensity and hold until any initial oscillation damps out. We then ramp down the intensity of the X-A$(v=0,v^\prime=0)$ transition over 15~ms and hold for 10~ms to ensure thermalization at the lower intensity. In the DC MOT, we find the temperature decreases with decreasing intensity from $T=5$~mK to $T=560(110)$ $\mu\text{K}$ at $I_0/64$. At lower intensities, the cloud size rapidly grows due to the limited restoring force. The RF MOT reaches lower temperatures. At $I_0/128$, $T=340(20)$ $\mu\text{K}$, close to the Doppler limit of $200$ $\mu\text{K}$ (Figure \ref{tempscan}). 

In order to estimate the number of molecules in the MOT, we measure the scattering rate by shuttering the X-A$(v=2,v^\prime=1)$ repump and monitoring the decay of fluorescence as the molecules are pumped into $v=2$. With a MOT beam intensity of $I_0$, we find a scattering rate of $1.7(1) \times 10^6$~s$^{-1}$ and determine a peak MOT number of $7(2) \times 10^4$ molecules for the DC MOT and $N= 1.1(3) \times 10^5$ molecules in the RF MOT, which corresponds to a peak density of $n_0=4(1) \times 10^6$~cm$^{-3}$, saturating with field gradient around 12~G/cm RMS. The density of trapped molecules is an order of magnitude greater than that reported for other molecular MOTs. Although further exploration is needed to confirm the exact factors that contribute to this improvement, there are indications that the slow, cold beam source plays a role. In particular, the distance between the source cell and the MOT chamber is less than half that of other molecular MOTs. This lowers the amount of molecular pluming during the slowing process and generally increases the solid angle for molecules that can be caught in the MOT. 



We have demonstrated both RF and DC magneto-optical trapping of CaF with $1.1(3) \times 10^5$ molecules trapped at a density of $n_0=4(1) \times 10^6$~cm$^{-3}$. We reach a temperature of  $340(20)$~$\mu\text{K}$ in the RF MOT. A direct route towards increasing this number further is possible using either a magnetic lens or transverse cooling at the output of the beam source to increase the CaF slow molecular beam intensity. We have realized high enough densities to load into optical tweezers and create optical arrays~\cite{lukin16array,barredo16array} for quantum simulation. With the even higher densities expected from the possible improvements, one could load an optical trap with large numbers for further evaporative cooling (or sympathetic cooling). Promising candidates for such cooling have been identified~\cite{lim15}, with a high ratio of elastic to inelastic collisions. With further evaporative cooling, a path exists towards creating a BEC of CaF.

This work was supported by the ARO, the AFOSR, and the NSF. BLA acknowledges support from the NSF GRFP. We would like to thank Dave DeMille and Michael Tarbutt for helpful discussions and the Tarbutt group for confirming the $v=3$ repump frequency.


\bibliography{motbib} 
\end{document}